\documentclass[final,5p,times,twocolumn]{elsarticle}
\usepackage{graphicx} 
\usepackage{amsmath,amssymb,amsfonts,amsthm}
\usepackage{hyperref}
\usepackage{xcolor}
\usepackage{geometry}
\usepackage{booktabs}
\usepackage{microtype}
\usepackage{enumitem}
\usepackage[T1]{fontenc}
\usepackage{lmodern}
\usepackage{tikz}
\usepackage{bm}
\usepackage[compat=1.1.0]{tikz-feynman}
\usepackage{ulem}
\usepackage{multirow}

\hypersetup{
  colorlinks=true,
  linkcolor=blue!70!black,
  citecolor=red!70!black,
  urlcolor=teal
}

\newcommand{\mHH}{m_{HH}}

\newcommand{\MSbar}{\overline{\rm MS}}

\usepackage[most]{tcolorbox}
\usepackage{xcolor}

\usepackage{physics}          
\usepackage{hyperref}
\usepackage{xcolor}
\usepackage{booktabs}
\usepackage{geometry}
\usepackage{enumitem}
\biboptions{sort&compress}

\hypersetup{colorlinks=true,linkcolor=blue,citecolor=blue,urlcolor=blue}

\theoremstyle{remark}

\journal{Physics Letters B}

\begin{document}

\begin{frontmatter}

\title{Top-Quark Mass Scheme Uncertainties in Higgs Boson Pair Production via Gluon Fusion}

\author[first]{Clara Del Pio}
\affiliation[first]{organization={Physics Division, 
    Brookhaven National Laboratory},
            city={Upton},
            postcode={11973}, 
            state={NY},
            country={USA}}

\author[second]{Sebastian Jaskiewicz}  
\affiliation[second]{organization={Albert Einstein Center for Fundamental Physics, Institut f\"ur Theoretische Physik},
            addressline={Universit\"at Bern}, 
            city={Bern},
            postcode={3012}, 
            country={Switzerland}}

\author[second]{Rebecca von Kuk}

\author[first]{Robert Szafron}

\begin{abstract}
We investigate theoretical uncertainties arising from the choice of  the top-quark mass scheme in Higgs boson pair production via gluon fusion. 
First, we implement the resummation of the top-quark mass logarithms obtained for the $gg\to HH$ amplitude at high energy and study its phenomenological impact on cross sections. Second, we formulate a threshold-aware treatment of the top-quark mass based on short-distance mass definitions, combining an MSR description near the top-pair threshold with an $\overline{\mathrm{MS}}$ treatment in the high-energy regime. 
Within the dynamical prescription, the combined mass-scheme variation
of the SM inclusive NLO cross section at $\sqrt{s}=14$~TeV is $^{+5.2\%}_{-0.2\%}$.
This estimate provides an alternative to the conventional
OS--$\overline{\mathrm{MS}}$ envelope.
\end{abstract}

\end{frontmatter}

\section{Introduction}

A major goal of the High-Luminosity LHC programme is the measurement of Higgs boson pair production~\cite{Cepeda:2019klc}.
In the Standard Model (SM), Higgs boson pair production via gluon fusion,
$gg \to HH$~\cite{Eboli:1987dy,Glover:1987nx,Plehn:1996wb},
provides direct sensitivity to the Higgs boson self-coupling, which is a key parameter of the electroweak symmetry-breaking sector.

A precise extraction of the Higgs self-interaction requires accurate theoretical predictions, which have improved substantially over the last decade. 
In the heavy-top limit, the QCD corrections are known at NLO~\cite{Dawson:1998py} and NNLO~\cite{deFlorian:2013uza,deFlorian:2013jea,Grigo:2014jma}, and the results are available up to $\mathrm{N}^3\mathrm{LO}$ and $\mathrm{N}^3\mathrm{LO}+\mathrm{N}^3\mathrm{LL}$~\cite{Banerjee:2018lfq,Chen:2019lzz,Chen:2019fhs,Chen:2026zmi,AH:2022elh}. 
Exact top-quark mass effects are known at NLO QCD~\cite{Borowka:2016ehy,Borowka:2016ypz,Baglio:2018lrj,Baglio:2020ini} and matched to parton showers~\cite{Heinrich:2017kxx,Jones:2017giv,Bagnaschi:2023rbx}.
At NNLO, approximate predictions including finite-mass effects are available~\cite{Grazzini:2018bsd}, also matched to parton showers~\cite{Alioli:2025xcu,Garosi:2026jyo}.
Moreover, various limits have been studied: the high-energy limit $s, |t|, |u| \gg m_t^2 \gg m_H^2$~\cite{Davies:2018ood,Davies:2018qvx,Davies:2019dfy,Davies:2023vmj},
the small Higgs boson mass limit: $s, |t|, |u| \gg m_H^2$~\cite{Xu:2018eos,Wang:2020nnr},
small-$p_T$ and/or small-$t$~\cite{Bonciani:2018omm,Bellafronte:2022jmo,Davies:2023vmj},
and expansion near the top-quark threshold: $s \sim 4 m_t^2$~\cite{Grober:2017uho}.
Electroweak corrections have also recently been computed~\cite{Muhlleitner:2022ijf,Davies:2022ram,Bi:2023bnq,Davies:2023npk,Heinrich:2024dnz,Bhattacharya:2025wnr,Davies:2025wke,Bonetti:2025vfd,Davies:2026wbx}, with the complete result given in Ref.~\cite{Bi:2023bnq}.
The current LHC Higgs Working Group recommendation~\cite{H:2026ksi} combines the available higher-order QCD and electroweak corrections into state-of-the-art predictions for the inclusive cross section and the $m_{HH}$ distribution.

Despite these advances, the treatment of the top-quark mass remains a significant source of theoretical uncertainty~\cite{Baglio:2020wgt}, as it is in other loop-induced processes such as off-shell Higgs and $ZH$ production~\cite{Amoroso:2020lgh,Mazzitelli:2022scc,Chen:2022rua,Degrassi:2022mro}.
The present recommendation~\cite{H:2026ksi} quotes a sizeable uncertainty arising from the top-mass scheme and scale variations. The prescription is based on comparisons between the cross-sections obtained with on-shell and $\overline{\mathrm{MS}}$ values of the top mass~\cite{Baglio:2020wgt}.
The on-shell (OS) prediction is taken as the central value. The uncertainty is then defined as the bin-by-bin envelope of predictions with $\overline{\mathrm{MS}}$ mass $\overline{m}_t(\mu_t)$ at $\mu_t=\overline{m}_t$ and $\mu_t=m_{HH}/4,\, m_{HH}/2,\,m_{HH}$, integrated over Higgs-pair invariant mass $m_{HH}$~\cite{Baglio:2018lrj,Baglio:2020ini,Baglio:2020wgt}.
Numerically, it reaches the level of
${}^{+5\%}_{-16\%}\, ,$ for inclusive cross sections at $\sqrt{s}=14~\mathrm{TeV}$, and is
significantly larger than the residual perturbative scale variation~\cite{H:2026ksi}.
The uncertainty due the top-mass-scheme choice is even more pronounced in differential distributions, both just below the $t\bar t$ threshold and in the high-energy tail: at NLO it reaches $-34\%$ at $m_{HH}=300$~GeV and $-35\%$ at $m_{HH}=1.2$~TeV~\cite{Baglio:2018lrj}.

This behaviour reflects the multi-scale nature of the process. Higgs-pair production spans kinematic regimes ranging from the  top-pair threshold region,
\begin{equation}
\label{eq:threshold_reg}
m_{HH} \sim 2m_t\, ,
\end{equation}
to the high-energy regime 
\begin{equation}
m_{HH} \gg m_t\, .
\end{equation}
Consequently, a single choice of a renormalisation scheme and scale cannot minimise higher-order corrections throughout the entire $m_{HH}$ range.
In particular, the commonly adopted choice $\mu_t \sim m_{HH}$ for the running top-quark mass leads to unnaturally large higher-order corrections near the top threshold, where the virtual top quarks are nearly on-shell and the relevant non-relativistic energy scale is set by
$\Delta = m_{HH} - 2m_t^{\mathrm{OS}}$. 
Even virtual corrections retain sensitivity to threshold effects through Coulomb singularities~\cite{Melnikov:1994jb,Kawabata:2016aya,Chen:2019fla}, albeit less pronounced than in real top-pair production.
As in top-pair production near threshold, this motivates the use of short-distance mass schemes adapted to the kinematic regime under consideration~\cite{Beneke:1998rk,Hoang:1998ng,Hoang:1999zc,Hoang:2008yj,Hoang:2017suc,Hoang:2020iah,Makela:2023xnt}.
Close to the di-Higgs threshold, where the heavy-top limit can be taken~\cite{Baglio:2020wgt}, $m_t$ sets the matching scale rather than appearing as a dynamical running parameter.  At large $m_{HH}$, the high-energy expansion instead favours $\mu_t\sim m_{HH}$~\cite{Baglio:2020ini,Davies:2018qvx}.

Physical predictions are scheme independent to all orders.  The residual scheme dependence at fixed order therefore reflects missing higher-order terms, and it decreases with the perturbative order.
Indeed, first three-loop results for the box form factor in the forward and large-$N_c$ limits show a reduced difference between the OS and $\overline{\mathrm{MS}}$ predictions at NNLO~\cite{Davies:2025ghl}, building on Refs.~\cite{Davies:2023obx,Davies:2024znp}.

In the recommendation~\cite{H:2026ksi}, the OS scheme is used for the central value, not least because the most precise predictions, such as NNLO$_{\rm FTapprox}$~\cite{Grazzini:2018bsd}, are currently only available in this scheme.
In this situation, the envelope of OS and $\overline{\mathrm{MS}}$ predictions offered a conservative estimate of the scheme dependence~\cite{Baglio:2020wgt}, which can now be refined using the improved understanding of the underlying higher-order structure.

In particular, a substantial part of this envelope reflects known higher-order information. Studies of the high-energy structure of the $gg\to HH$ amplitude have clarified the origin of the large logarithms driving the discrepancy between on-shell and $\overline{\mathrm{MS}}$ predictions~\cite{Jaskiewicz:2024xkd}. In the regime
$ s,|t|,|u| \gg m_t^2 \gg m_H^2$,
the amplitudes develop logarithms of the form
$\ln\frac{m_t^2}{s}\, ,$
whose leading structure is controlled by top-mass renormalisation at leading power (LP) in $m_t^2/s$.
In the $\overline{\mathrm{MS}}$ scheme with $\mu_t\sim\sqrt{s}$ these logarithms are resummed implicitly by the running mass, whereas the OS result, obtained with a scheme conversion truncated at $\mathcal{O}(\alpha_s)$, contains only the first of them.
The OS--$\overline{\mathrm{MS}}$ difference at high energy is therefore driven largely by the truncation of a tower of logarithms that is known to all orders. 
Reinstating this tower in the OS amplitude substantially reduces the scheme dependence at high energies~\cite{Jaskiewicz:2024xkd}.
On this basis, Ref.~\cite{Jaskiewicz:2024xkd} argued that such a difference should not be counted as a theoretical uncertainty: known contributions belong in the central prediction, and only genuine unknowns in the uncertainty budget.
Two questions were left open~\cite{Jaskiewicz:2024xkd,H:2026ksi}: the impact at the level of the cross section, including the real-emission phase space, and the threshold region $300~\mathrm{GeV}\lesssim m_{HH}\lesssim 800~\mathrm{GeV}$, which dominates the total rate but is outside the high-energy regime.
We address both questions and propose a physically motivated estimate of the top-mass scheme uncertainty for inclusive $gg\to HH$ cross sections.

First, we implement the leading-logarithmic high-energy resummation~\cite{Jaskiewicz:2024xkd} in predictions for the $m_{HH}$ spectrum, switching it on smoothly with a profile function~\cite{Ligeti:2008ac,Abbate:2010xh,Lustermans:2019plv}.
Since the resummed OS prediction approaches the $\overline{\mathrm{MS}}$ one at high $m_{HH}$, the remaining ambiguity is the location of the transition, which we parametrise by a single variable so that experimental fits can treat it as a nuisance parameter~\cite{Cal:2023mib}.
However, below the onset of the resummation, this approach retains the OS prediction without a scheme uncertainty.
This limitation motivates our second approach.

Second, we construct a threshold-aware prescription based on the MSR mass scheme, in which the top mass freezes below the threshold region and matches onto the $\overline{\mathrm{MS}}$ mass at higher scales, continuously for the default transition point. 
This construction incorporates the hierarchy of physical scales relevant to Higgs pair production and avoids extending the high-scale running into the top-threshold region defined in Eq.~\eqref{eq:threshold_reg}.
The resulting description provides a systematic EFT-inspired interpolation between threshold and high-energy dynamics, together with its own uncertainty from variations of $\mu_t$ and of the MSR--$\overline{\mathrm{MS}}$ transition point.
We compare this approach with previous analyses of top-mass scheme uncertainties in Higgs pair production~\cite{Baglio:2018lrj,Baglio:2020ini,Baglio:2020wgt,Bagnaschi:2023rbx}.

We finally note that our analysis does not include Coulomb resummation effects~\cite{Fadin:1987wz,Strassler:1990nw,Hoang:1999zc} near threshold. Such contributions are expected to become relevant in the region $\Delta \ll m_t$ and may eventually need to be combined consistently with the framework developed here. 
However, our primary focus is the dominant uncertainty associated with top-mass renormalisation and logarithmic scale hierarchies. Their consistent combination with the present framework is left for future work.

\section{Resummation of high-energy logarithms}
\label{sec:resummation}
Ref.~\cite{Jaskiewicz:2024xkd} investigated the structure of the $gg\to HH$ amplitude in the high-energy limit and showed that, at leading power in the $m_t^2/s$ expansion, the full tower of leading large logarithms $\ln\frac{m_t^2}{s}$ appearing in the OS amplitude can be predicted from the renormalisation of the top-quark mass.
The truncation of this tower of logarithms in fixed-order calculations was identified as the driving source of the discrepancy between predictions in the on-shell and $\overline{\mathrm{MS}}$ $m_t$-schemes.
Once this tower of logarithms is reinstated, the agreement between the two predictions in different schemes improves drastically.
Since the leading logarithms (and in principle beyond) in the conversion factor are \textit{known}, the difference between the two schemes due to a mismatch in the truncation order should not be part of the uncertainty budget. 
At the amplitude level, the top-mass scheme uncertainty given by comparing the OS and central $\overline{\mathrm{MS}}$ results is reduced from 65\% to 18\% at LO and from 36\% to 11\% at NLO~at~8~TeV. 

The remaining open question is the impact of this result on the mass-scheme uncertainty for the $gg\to HH$ cross section.
To address this, we implement the results of~\cite{Jaskiewicz:2024xkd} by modifying the code \texttt{ggxy}~\cite{Davies:2025qjr}, which combines the high-energy~\cite{Davies:2018ood,Davies:2018qvx, Davies:2019dfy} and forward~\cite{Davies:2023vmj} expansions of the two-loop virtual amplitude.
In particular, we modify the form factors making up the $gg\to HH$ amplitude according to equations~(4.23) and~(4.24) of~\cite{Jaskiewicz:2024xkd}.
The amplitude is written in terms of the two helicity amplitudes, separated into box ($y_t^2$) and triangle ($y_t\lambda_3$) contributions and expanded as
\begin{align}
\label{eq:amp-def}
A_{i,j} = \sum_{k\geq0} \left(\frac{\alpha_s}{2\pi}\right)^{\!k} A^{(k)}_{i,j},
\end{align}
where $i=1,2$, and $j = y_t^2,\,y_t\lambda_3$. 
The triangle contributions are power-suppressed by $m_H^2/s$ relative to the box ones and are not modified here.
The resummed scalar  amplitudes are given by~\cite{Jaskiewicz:2024xkd} 
\begin{align}
\label{eq:a_ll}
A^{(k,\,\text{LL})}_{i,y_t^2}(m_t^\mathrm{OS}) =
\left(\frac{m^\text{LL}(\mu_t)}{m_t^\mathrm{OS}} \right)^2
A^{(k)}_{i,y_t^2}(m_t^\mathrm{OS}), 
\end{align}
where 
\begin{align}\label{eq:schmeme-conv-all-order}
m^\text{LL}(\mu) = m_t^\mathrm{OS} \exp \left[ a_{\gamma_m}^\text{LL}(\mu)\right]\, z_m(m_t^\mathrm{OS}),
\end{align}
and in turn
\begin{align}
a_{\gamma_m}^\text{LL}(\mu) = \frac{3 C_F}{2 \beta_0} \ln \left(1 - \frac{\alpha_s(\mu)}{2 \pi} \beta_0 \ln \left(\frac{\mu^2}{{m_t^\mathrm{OS}}^2} \right) \right).
\end{align}
We use the tree-level result for $z_m$, which is the ratio of the OS and $\overline{\mathrm{MS}}$ renormalisation factors known to four loops~\cite{Chetyrkin:1999ys, Melnikov:2000qh,Marquard:2015qpa,Marquard:2016dcn}.
$\mu_t$ is the renormalisation scale for the top-quark mass in the $\overline{\mathrm{MS}}$ scheme and $\beta_0 = 11 C_A/6 - 2/3 T_F n_f$. 
The leading-power resummation applies at high invariant mass but does not provide a controlled approximation throughout the spectrum. We switch it on smoothly using a profile function $f(\mHH,\{x_i\})$, following
Refs.~\cite{Ligeti:2008ac,Abbate:2010xh,Lustermans:2019plv}.
The parameters $\{x_i\}$ specify the transition region; we abbreviate the profile as $f$ below.

\begin{figure}
    \centering
    \includegraphics[width=\linewidth]{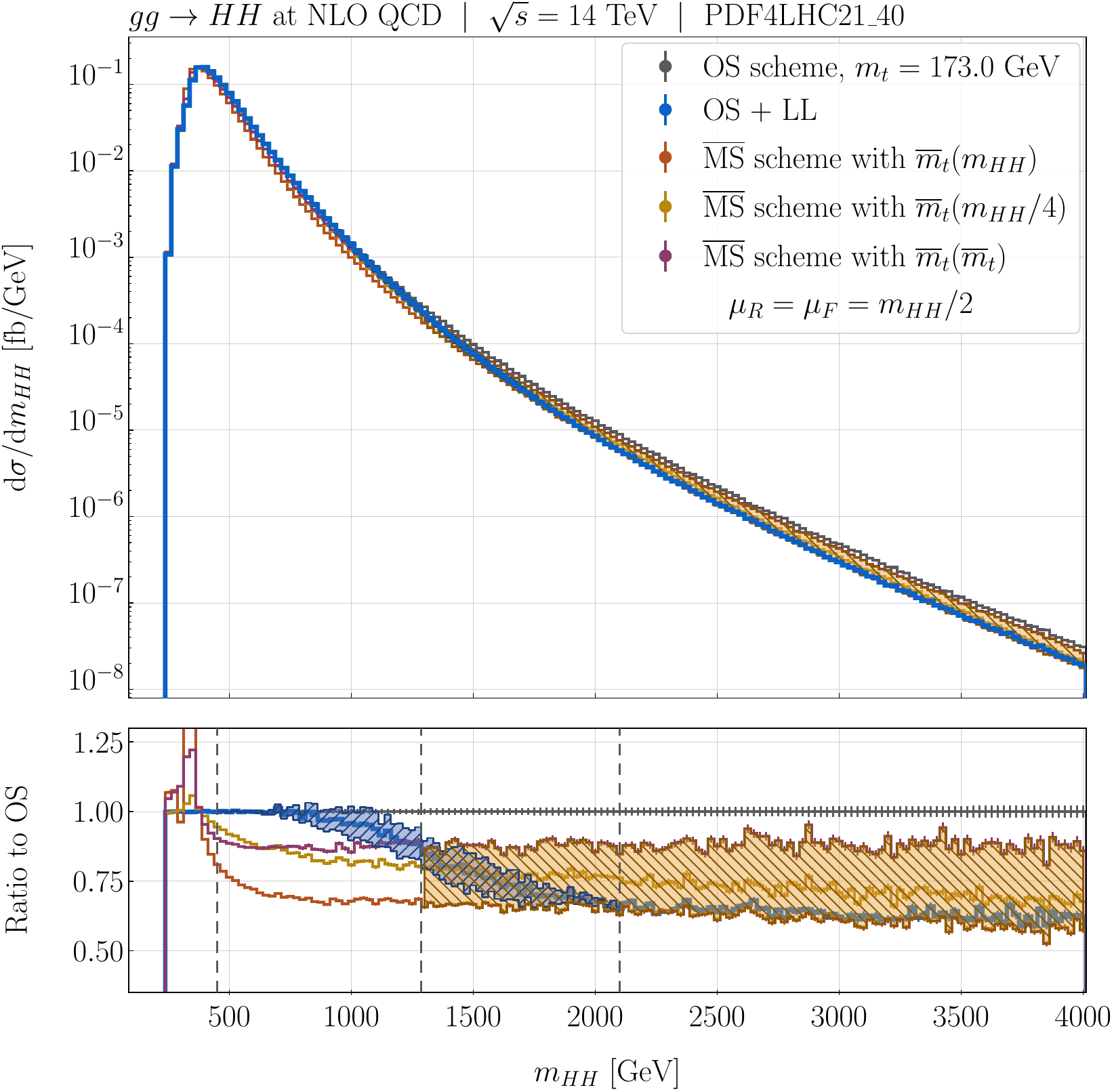}
    \caption{
    Cross section prediction at $\sqrt{s}=14$~TeV, as a function of the di-Higgs invariant mass $m_{HH}$ turning on the resummation of high-energy logs using profile scales.
    The blue envelope corresponds to the matching uncertainty, whereas the orange envelope shows the $\mu_t$ variations.
    Parton distributions are evaluated using the PDF4LHC21\_40~\cite{PDF4LHCWorkingGroup:2022cjn} set through LHAPDF~\cite{Buckley:2014ana}, with $\alpha_s(m_Z)=0.118$.
    Scale variations are omitted for better readability.
    }
    \label{fig:HE_resummation}
\end{figure}

As shown in Fig.~\ref{fig:HE_resummation}, the resummed OS prediction approaches the  $\overline{\mathrm{MS}}$ prediction with $\mu_t=m_{HH}$ at high invariant mass. The profile controls the transition from the fixed-order OS result to this high-energy prediction.
We vary the profile to estimate the associated matching uncertainty.

We apply the resummation of Ref.~\cite{Jaskiewicz:2024xkd} only to the leading-power amplitude, writing $A^{(k)}_{i,y_t^2}=A^{(k),\mathrm{LP}}_{i,y_t^2}+\mathcal{O}(y_t^2m_t^2/s)$.
We keep the power-suppressed terms at fixed order because the leading-power resummation used here does not determine their logarithms.
Then, the Born contribution is evaluated with
\begin{align}
\label{eq:profile}
A^{(0)}_{i,y_t^2} &\to A^{(0)}_{i,y_t^2}
+ f\,\Bigl[A^{(0,\,\text{LL}),\mathrm{LP}}_{i,y_t^2} - A^{(0),\mathrm{LP}}_{i,y_t^2}\Bigr],
\end{align}
and, to avoid double counting, when the code is run in \texttt{nlo} mode, we also remove the higher-order term from the resummed exponential in the matching
\begin{align}
\label{eq:profile-sub}
A^{(0)}_{i,y_t^2} &\to A^{(0)}_{i,y_t^2}
+ f\,\Bigl[A^{(0,\,\text{LL}),\mathrm{LP}}_{i,y_t^2} - A^{(0),\mathrm{LP}}_{i,y_t^2}\Bigr]
\notag\\
&\quad - 2f\,a_{\gamma_m}^\text{LL}(\mu_t)\,A^{(0),\mathrm{LP}}_{i,y_t^2},
\end{align}
where $2a_{\gamma_m}^\text{LL}$ is the exponent of $(m^\text{LL}/m_t^\mathrm{OS})^2$ in Eq.~\eqref{eq:schmeme-conv-all-order}.
The real-emission contribution is evaluated with \texttt{Recola}~\cite{Actis:2012qn,Actis:2016mpe} in \texttt{ggxy}, in the OS scheme, and with the unmodified Born amplitude.
We checked that for the $m_{HH}$ distribution the effect of mass scheme change in the real correction is negligible.\footnote{Real radiation has a significant effect for non-inclusive observables such as $p_T$ or jet veto.}

The transition starts at $\mHH = x_1$ and ends at $\mHH = x_3$, with $x_2$ as the turning point.
Following Ref.~\cite{Cal:2023mib}, we fix $x_1$ and $x_3$ to their lowest and highest reasonable values, $450~\mathrm{GeV}$ and $2.1~\mathrm{TeV}$, and vary only $x_2$.
The matching uncertainty is then described by a single parameter, which can be treated as a nuisance parameter and simplifies error propagation in experimental analyses.
Our central choice is $x_2 = (x_1+x_3)/2 = 1.275~\mathrm{TeV}$, and we vary $x_2$ between $0.6$ and $1.8~\mathrm{TeV}$.
The matching uncertainty $\Delta_{\rm match}$ is given by the maximum envelope of the resulting predictions.
The results are shown in Fig.~\ref{fig:HE_resummation}, where the matching uncertainty corresponds to the blue band, which by construction is non-zero only in the transition region $[x_1,x_3]$.
Integrating over the full invariant-mass range gives the following uncertainty due to the switch on point for the resummation\footnote{We use the notation $\Delta \sigma$ to denote relative uncertainties on the cross section.}
\begin{align}
\label{eq:resummed_unc}
\Delta\sigma_{\rm match} =  \pm 0.25\% \, .
\end{align}
Thus far we have shown that the resummation of the LP leading logarithms reproduces the $\overline{\mathrm{MS}}$ results in the high-energy region. To do this, we have chosen to reproduce an $\overline{\mathrm{MS}}$ curve with a $\mu_t$ which is furthest away from the OS scheme giving rise to the large uncertainties in \cite{Baglio:2020ini}. 

The precise setting of the scale of the $\overline{\mathrm{MS}}$ mass is a subleading logarithmic effect, however given the fact that with a specific scale choice any of the $\overline{\mathrm{MS}}$ lines can be reproduced, we still include the spread of the predictions obtained with the $\overline{\mathrm{MS}}$ mass at different $\mu_t$ in the high-energy region as part of our uncertainty. This is captured by the orange band in Fig.~\ref{fig:HE_resummation} and yields the following uncertainty on the total cross section
\begin{align}
\label{eq:mutvary}
\Delta\sigma_{\mu_t} = \pm 0.023\%.
\end{align}
As the matching and the $\mu_t$ variation probe two conceptually different sources of uncertainty, we treat them as independent and add them in quadrature.
In total, the top-quark mass scheme uncertainty of this prescription amounts to
\begin{align}
\label{eq:resummed_unc+mutvary}
    \Delta\sigma_{m_t} =\sqrt{\Delta\sigma_{\rm match}^2+\Delta \sigma_{\mu_t}^2}\,
=  \pm 0.25\%
\end{align}
to be compared with the ${}^{+5\%}_{-16\%}$ of the fixed-order envelope~\cite{H:2026ksi}, which, unlike Eq.~\eqref{eq:resummed_unc+mutvary}, also varies the region $m_{HH}<x_1$.

The smallness of Eq.~\eqref{eq:resummed_unc+mutvary} should, however, be interpreted with care, as it rests on two simplifications.
First, the resummation is applied only to the virtual $gg\to HH$ amplitude, since the logarithmic structure has so far been established only for the $2\to2$ process.
The real-emission contributions are therefore still evaluated in the OS scheme at fixed order.
Although their numerical impact is small, treating the top mass differently in the virtual and real contributions is not fully consistent.

The next section introduces mass-scheme variations just above the
top-pair threshold, while retaining the pole mass below threshold.
The OS scheme is well motivated in the threshold region, as it places the $t\bar t$ branch point at its physical location.
Nevertheless, it remains a choice, and other short-distance mass definitions are equally suited to this region.
Since the bulk of the cross section lies at $m_{HH}\lesssim x_1$, this assumption largely determines the size of Eq.~\eqref{eq:resummed_unc+mutvary}.
In the following section, we develop a treatment of the top mass that provides a scheme uncertainty in the threshold region as well.

\section{Threshold-aware top-mass schemes}
\label{sec:msr}

For a stable top quark, the $gg\to HH$ amplitude develops a
$t\bar t$ branch point at $m_{HH}=2m_t^{\mathrm{OS}}$.
Its discontinuity is related by unitarity to an on-shell $t\bar t$
intermediate state produced by the gluons and annihilating into $HH$.
The $\MSbar$ scheme displaces this branch point from its physical location, generating large corrections near threshold that are artefacts of the scheme choice. A pole-mass-based scheme, such as the on-shell scheme, is preferred for the top mass appearing in the loops.\footnote{The same scheme is used for the top mass in the Yukawa couplings. Near threshold the amplitude factorises into hard vertices and a non-relativistic part~\cite{Kawabata:2016aya,Chen:2019fla}. The Yukawa couplings enter the hard coefficients of the factorised  amplitude, where a short-distance mass at $\mu\sim 2m_t$ is natural. The difference does not induce large logarithmic corrections.}

Using the $\overline{\rm MS}$ mass at high $m_{HH}$ and the pole mass below $m_{HH} \sim 2 m_t^\mathrm{OS}$ would, however, cause the $m_{HH}$ distribution to present a jump, as the top-quark mass, and with it the form factors, would change discontinuously at the switching point.
We thus adopt an $\mHH$-dependent mass scheme, that uses the MSR mass in the transition region and is continuous for the default transition point.

The MSR mass scheme is a low-scale short-distance mass  designed for non-relativistic heavy-quark dynamics, where fluctuations at the scale of the mass are integrated out~\cite{Hoang:2008yj,Hoang:2014oea,Hoang:2017suc}. 
It extends the $\MSbar$ running to scales $R$ below the mass, with an evolution that is linear in $R$, and thus interpolates between the $\MSbar$ and OS schemes: for $R\to0$ it reduces to the pole mass, up to the $\mathcal{O}(\Lambda_{\rm QCD})$ renormalon ambiguity~\cite{Beneke:1994sw,Bigi:1994em,Hoang:2017suc}.

For di-Higgs production, the natural choice to set the starting point of the evolution is $m_{HH} = 2 m_t^{\rm OS}$, i.e. the $t \bar{t}$ threshold. 
We therefore adopt the practical definition of the MSR mass~\cite{Hoang:2017suc}
\begin{equation}
    \label{eq:MRS}
    m_t^\mathrm{OS} - m_t^{\rm MSR}(R) = R \sum_{n=1}^4 a_n(n_\ell) \Big( \frac{\alpha_s^{n_\ell}(R)}{4 \pi}\Big)^n \, ,
\end{equation}
with $R=\Delta=m_{HH}-2 m_t^{\rm OS}$. This ensures that the MSR mass reduces to OS mass at the threshold, where $R=0$.
In Eq.~\eqref{eq:MRS} the number of light flavours is set to $n_\ell = 5$ and the $a_n(n_\ell)$ coefficients, up to $n=4$, are given in~\cite{Hoang:2017suc}. 
We implement a new ``dynamical'' mass scheme in our version of \texttt{ggxy}, motivated by the fixed-$R$ MSR analysis of the $t\bar t$ invariant-mass distribution in~\cite{Makela:2023xnt}. In our scheme
\begin{equation}
   \label{eq:dynamic_mass}
    m_t(R) = \begin{cases} 
     m_t^\mathrm{OS} \quad & R \le 0 \\
     m_t^{\rm MSR}(R) \quad &  0 < R\le \delta\\
     \overline{m}_t(R) \quad &  R > \delta \, ,
\end{cases}
\end{equation}
and $m_t^{\rm MSR}(R)$ is defined as in Eq.~\eqref{eq:MRS}. Note that the scale $R$ for the running of the top mass is different from the renormalisation and the factorisation scales, $\mu_R$ and $\mu_F$, used in the amplitude calculation\footnote{A floor of $2$ GeV has been implemented for the $\alpha_s(R)$ coupling entering the MSR top mass, to prevent it from entering the non-perturbative regime.}. For high values of $R$, we use the default {\tt ggxy} implementation of the $\MSbar$ mass $\overline{m}_t (R)$~\cite{Davies:2025qjr}. 
To ensure continuity of the top mass, the value of the transition parameter $\delta$ between the MSR and $\MSbar$ masses should be set to $\delta = \overline{m}_t (\overline{m}_t) \simeq 162$~GeV, because of the matching condition $m_t^{\rm MSR}(m_t^{\rm MSR}) = \overline{m}_t (\overline{m}_t)$. We take the variation in the value of $\delta$ as an additional uncertainty for this scheme implementation, and check that for phenomenologically relevant observables the variation of $\delta$ (that would make $m_t(R)$ discontinuous) is subleading to the $R$ variation in the total error budget\footnote{We present here the results for the $m_{HH}$ distribution. We checked that the error due to the variation of $\delta$ is subleading for all distributions produced in {\tt ggxy}, including distributions in $p_T$.}. 

Following the notation of~\cite{Davies:2023vmj,Davies:2025qjr}, the virtual corrections to $gg \to HH$  can be written in terms of the form factors $F_i = \sum_{k\geq0} \left( \frac{\alpha_s(\mu_R)}{\pi} \right)^k F_i^{(k)}$ with $i=1,2$ and $k$ denoting the power of $\alpha_s$ relative to the LO. In our dynamical scheme, the two-loop form factors are modified as follows
\begin{equation}
    F_{i, \, \rm dyn. \; scheme}^{(1)} = F_{i, \rm OS}^{(1)} + 2 \, C_1 \, m_t^2 \, \frac{\partial F_{i, \rm OS}^{(0)}(m_t^2)}{\partial m_t^2} \, ,
\end{equation}
with $i=1,2$ and $C_1$ defined as
\begin{equation}
    C_1 = \begin{cases} 
     0 \quad & R \le 0 \\
     C_F \frac{R}{m_t^{\rm MSR}(R)}  \quad & 0 < R \le \delta\\
     C_F \, \left( 1+\frac{3}{4} \ln \frac{R^2}{\overline{m}_t^2(R)} \right)  \quad & R > \delta \, .
     \end{cases}
\end{equation} 
The real matrix elements are consistently treated by passing the top mass in Eq.~\eqref{eq:dynamic_mass} to {\tt Recola} \cite{ Actis:2012qn,Actis:2016mpe}. This ensures the correct implementation of the full NLO calculation in the dynamical scheme choice presented here.
\begin{figure}
    \centering
    \includegraphics[width=\linewidth]{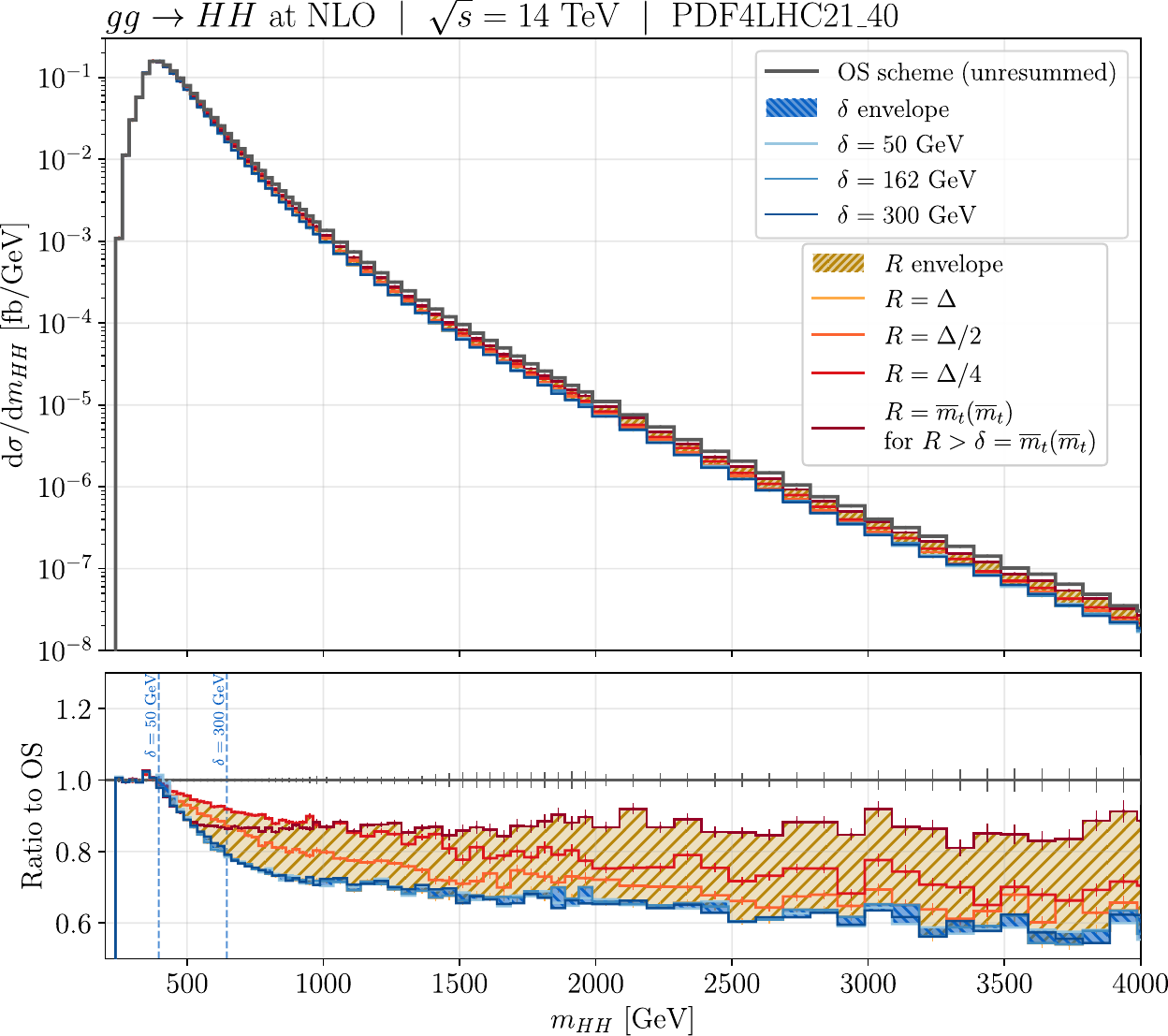}
    \caption{Upper panel: cross section prediction at $\sqrt{s}=14$~TeV, as a function of the di-Higgs invariant mass $m_{HH}$, in the OS scheme and the variants of the dynamical scheme discussed in the text. Lower panel: ratio of the predictions to the OS curve. The blue envelope corresponds to the uncertainty due to the choice of the transition point $\delta$, while the brown one shows the variation of the top-mass scale $R$. Monte Carlo errors are shown as bars. We do not display renormalisation and factorisation scale variations.
    }
    \label{fig:MSR}
\end{figure}

We consider two sources of uncertainty in this implementation: one stemming from scale variations in $R$ and the other due to the choice of the transition point $\delta$. The former is quantified by varying the scale $R= \Delta$ by factors $1/2,1/4$, so that each realisation of the MSR scheme asymptotically tends to a $\MSbar$ top mass realisation with $m_{HH}/2$ or $m_{HH}/4$ , respectively, in the high-energy regime. A fourth prediction, corresponding to the $\MSbar$ top mass $\overline{m}_t (\overline{m}_t)$ for high invariant masses, is taken by using $\delta=\overline{m}_t(\overline{m}_t)$ and fixing $R=\overline{m}_t$ for $R>\delta$. 

In the $\mHH$ distribution, we take the bin-by-bin envelope (maximum and minimum) of these four predictions. The corresponding inclusive uncertainty is obtained by linearly integrating the bin-wise extrema with a simple rectangle rule.
The uncertainty due to the choice of $\delta$ is similarly estimated by taking the envelope of the predictions with $\delta=50, \, 162, \, 300$ GeV.

The resulting uncertainties for the $m_{HH}$ distribution are shown in Fig.~\ref{fig:MSR}. 
The blue envelope corresponds to the uncertainty due to the variation of $\delta$, which is smaller than that from $R$, while the uncertainty due to the scale variation $R$ is represented by the brown envelope. 
The bins in the high energy tail, approximately for $m_{HH} > 1500$~GeV, are weighted by a very small cross section and therefore do not contribute significantly to the total error estimation. 
With our statistics, in the high energy tail, the Monte Carlo noise becomes comparable with the estimated envelope in $\delta$.
A quantitative experimental analysis will require higher-statistics predictions in these bins.

On the total cross section, the final uncertainty associated with this dynamical scheme choice for the top mass is 
\begin{equation}
\Delta \sigma_{m_t}^{\rm LO} = {}^{+9.5\%}_{-0.3\%} (R) \; ^{+2.1\%}_{-0.1\%} (\delta) = {}^{+9.7\%}_{-0.3\%} ({\rm tot})\, ,
\end{equation}
at LO and 
\begin{equation}
\Delta \sigma_{m_t}^{\rm NLO} ={}^{+5.0\%}_{-0.2\%}(R) \; ^{+1.6\%}_{-0.1\%} (\delta)={}^{+5.2\%}_{-0.2\%} ({\rm tot})\, ,
\end{equation} at NLO, where the first number refers to the $R$ scale variation and the second to the choice of the transition point.
The total scheme uncertainty is obtained by assuming the two components to be uncorrelated and adding the upward and downward variations separately in quadrature.

\section{Impact on the LHC determination of the Higgs self-coupling}
\label{sec:pheno}

\begin{table*}[t]
\centering
\renewcommand{\arraystretch}{1.4}
\begin{tabular}{|c|c|c|c|c|c|c|c|c|c|}
\hline
\multicolumn{2}{|c|}{$\kappa_\lambda$} & $-5$ & $-1$ & $0$ & $1$ & $2$ & $2.4$ & $3$ & $5$ \\
\hline
\multicolumn{2}{|c|}{$\sqrt{s} = 13.6$~TeV $\sigma$ [fb]} & 539.3 & 118.6 & 63.37 & 28.06 & 12.69 & 12.13 & 17.28 & 86.32 \\
\hline
\multicolumn{2}{|c|}{Scale unc.} & ${}^{+17.2\%}_{-14.2\%}$ & ${}^{+16.4\%}_{-13.8\%}$ & ${}^{+15.8\%}_{-13.6\%}$ & ${}^{+15.0\%}_{-13.3\%}$ & ${}^{+15.3\%}_{-13.5\%}$ & ${}^{+16.7\%}_{-14.1\%}$ & ${}^{+18.7\%}_{-14.9\%}$ & ${}^{+19.1\%}_{-15.1\%}$ \\
\hline
 & $\Delta\sigma_{R}$ & ${}^{+1.3\%}_{-0.2\%}$ & ${}^{+2.5\%}_{-0.2\%}$ & ${}^{+3.3\%}_{-0.2\%}$ & ${}^{+5.0\%}_{-0.2\%}$ & ${}^{+6.6\%}_{-0.2\%}$ & ${}^{+5.4\%}_{-0.2\%}$ & ${}^{+2.4\%}_{-0.4\%}$ & ${}^{+0.2\%}_{-0.6\%}$ \\
\cline{2-10}
$m_t$ unc. & $\Delta\sigma_{\delta}$ & ${}^{+0.6\%}_{-0.1\%}$ & ${}^{+0.9\%}_{-0.1\%}$ & ${}^{+1.2\%}_{-0.1\%}$ & ${}^{+1.5\%}_{-0.1\%}$ & ${}^{+1.5\%}_{-0.1\%}$ & ${}^{+0.9\%}_{-0.1\%}$ & ${}^{+0.2\%}_{-0.5\%}$ & ${}^{+0.1\%}_{-0.6\%}$ \\
\cline{2-10}
 & {\boldmath$\Delta\sigma_{\rm tot}$} & {\boldmath${}^{+1.5\%}_{-0.2\%}$} & {\boldmath${}^{+2.7\%}_{-0.3\%}$} & {\boldmath${}^{+3.5\%}_{-0.2\%}$} & {\boldmath${}^{+5.2\%}_{-0.2\%}$} & {\boldmath${}^{+6.8\%}_{-0.3\%}$} & {\boldmath${}^{+5.5\%}_{-0.2\%}$} & {\boldmath${}^{+2.5\%}_{-0.6\%}$} & {\boldmath${}^{+0.2\%}_{-0.9\%}$} \\
\hline
\end{tabular}
\caption{NLO $gg\to HH$ cross section at $\sqrt{s}=13.6$~TeV for different values of $\kappa_\lambda$. The numerical inputs are chosen as in~\cite{wg4wiki}, and we use the  PDF4LHC21\_40 set~\cite{PDF4LHCWorkingGroup:2022cjn}. Scale uncertainties are shown as 7-point variation of $\mu_R,\mu_F$ around $m_{HH}/2$. The top-mass scheme uncertainties are estimated from the envelope of variations of  $R$ and of $\delta$ in the dynamical scheme described in the text. We give the total mass scheme uncertainty by combining in quadrature $\Delta \sigma_{R}$ and $\Delta \sigma_{\delta}$.}
\label{tab:xs_kappa}
\end{table*}

The top-mass uncertainty enters directly into the constraints on the Higgs self-coupling at the LHC. In the unitary gauge, the SM Higgs potential reads
\begin{equation}
    V(H) = \frac{m_H^2}{2} H^2 + \lambda_3 v H^3 + \frac{\lambda_4}{4} H^4 \, .
\end{equation}
In the SM, the Higgs mass $m_H$ and the trilinear and quartic interactions, $\lambda_3$ and $\lambda_4$ respectively, can be related as
\begin{equation}
\lambda_3^{\rm SM} = \lambda_4^{\rm SM} = \frac{m_H^2}{2 v^2} \, ,
\end{equation}
with $v$ the vacuum expectation value of the Higgs field.
Double Higgs production via gluon fusion represents the main channel to probe the Higgs trilinear coupling. Although Higgs-pair production has not yet been observed due to the small cross section of this process, constraints on $\lambda_3$ can be derived in terms of a possible deviation from its SM value,  parametrised as
\begin{equation}
    \kappa_\lambda = \frac{\lambda_3}{\lambda_3^{\rm SM}} \, .
\end{equation}
The present experimental limit is $-0.71 < \kappa_\lambda < 6.1$ at $95\%$ confidence level, 
from the combination of the ATLAS and CMS Run~2 searches~\cite{CMS:2026nuu}.

Since the leading-order $gg \to HH$ amplitude is the sum of triangle diagrams, which depend on $\kappa_\lambda$, and box diagrams, which do not depend on the Higgs self-coupling, the cross section has the quadratic form
\begin{equation}
\label{eq:kappapar}
\sigma(\kappa_\lambda) = A + B \kappa_\lambda + C \kappa_\lambda^2 \, .
\end{equation}
At LO, $A$ comes from the box diagrams, $C$ from the triangle diagrams, and $B$ from their interference. 
To avoid generating Monte Carlo samples at every value of $\kappa_\lambda$, experiments commonly produce full‑simulation samples at a few $\kappa_\lambda$ points and then use Eq.~\eqref{eq:kappapar} to extrapolate to any other value.

The standard sample points are $\kappa_\lambda = 0, \, \pm 1, \, 2, \, 2.4, \, 3 , \, \pm 5$, where $ \kappa_\lambda =2.4$ is close to the minimum of the cross section, at which the box–-triangle interference is maximally destructive.  
Table~\ref{tab:xs_kappa} shows the NLO  cross sections at $\sqrt{s}=13.6$~TeV in the dynamical scheme of Sec.~\ref{sec:msr}, which describes the transition from the triangle-dominated threshold region to the box-dominated high-energy region.
The top and Higgs masses are set in accordance with the table in~\cite{wg4wiki}. 

The uncertainty stemming from the 7-point variation of $\mu_F$ and $\mu_R$ around $m_{HH}/2$ is about  $^{+17\%}_{-14\%}$, with little dependence on the Higgs trilinear coupling. This differs from~\cite{wg4wiki}, which uses approximate NNLO predictions with a reduced scale uncertainty~\cite{Amoroso:2020lgh, Grazzini:2018bsd}.

The uncertainty related to the top-mass definition is estimated as described in Sec.~\ref{sec:msr}: the variation of $R$ has an effect below $7\%$, while the uncertainty due to the transition point is at most $1.5\%$, for every $\kappa_\lambda$. For comparison, the top-mass uncertainty from~\cite{wg4wiki}, based on the conservative envelope among the OS and $\MSbar$ schemes, can reach the level of $20\%$. With our statistics, the Monte Carlo error on the total cross section is about $0.03\%$, while the noise error on the envelope estimate is below $0.2\%$.

Note that, by construction, the variations of $R$ and $\delta$ have an effect only above the $t \bar{t}$ threshold. The uncertainty is largest around $\kappa_\lambda=2 - 2.4$, where the box–triangle cancellation enhances any change in the effective top mass. At $\kappa_\lambda=\pm 5$, the triangle contribution is enhanced and most of the cross section lies below $2m_t^\mathrm{OS}$, with a smaller resulting uncertainty. The corresponding uncertainty on $\kappa_\lambda$ in an experimental analysis will thus depend on the cross section behaviour as a function of $m_{HH}$. This dependence is  better represented as a shape uncertainty in the $m_{HH}$ distribution, for which the differential results presented here can be used. 

\section{Conclusions}
\label{sec:conclusions}

We have developed an improved, physics-driven treatment of the top-quark mass uncertainty in Higgs boson pair production via gluon fusion, which has become one of the dominant theoretical uncertainties for this process~\cite{H:2026ksi}.
Our two complementary approaches, implemented in \texttt{ggxy}, target the high-energy and threshold regions of the Higgs-pair invariant-mass spectrum, respectively.

First, we have included the all-order resummation of the LP LLs in the $gg\to HH$ amplitude derived in Ref.~\cite{Jaskiewicz:2024xkd}.
As these logarithms are fully determined by the renormalisation of the top-quark mass, the difference they induce between the OS and $\overline{\mathrm{MS}}$ predictions at high invariant masses is a known effect rather than a genuine uncertainty.
Combining the resummed and OS predictions through a profile function, we obtain a residual scheme uncertainty of $\pm 0.25\%$ on the total cross section, including the matching uncertainty and the dependence on the scale of the $\overline{\mathrm{MS}}$ mass.

Second, we have addressed the threshold region, where a pure $\overline{\mathrm{MS}}$ description displaces the $t\bar t$ branch point from its physical location.
We have introduced a dynamical top-quark mass scheme that interpolates between the pole mass below threshold, the MSR mass in the transition region, and the $\overline{\mathrm{MS}}$ mass at high invariant masses, and implemented it consistently in both the virtual and real contributions at NLO.
Varying the scale $R$ and the transition parameter $\delta$, we obtain an uncertainty on the total cross section of $^{+5.2\%}_{-0.2\%}$ at NLO.
Both developments show that, once the top-quark mass scheme is adapted to the physical scales of the process, its impact on di-Higgs predictions is substantially smaller than suggested by the conventional envelope of OS and $\overline{\mathrm{MS}}$ results.

We apply the dynamical scheme to the NLO cross section for $-5\le\kappa_\lambda\le5$ which forms the theoretical input for constraining the trilinear Higgs coupling at the LHC.
With this choice, the top-mass scheme uncertainty is reduced from up to $\sim 20\%$ in the current recommendation~\cite{wg4wiki} to at most $6.6\%$ from the variation of $R$ and approximately $1.5\%$ from the variation of $\delta$.
Thus, it is subleading with respect to the NLO renormalisation and factorisation scale uncertainty of about ${}^{+17\%}_{-14\%}$, but comparable to the ${}^{+2.1\%}_{-4.9\%}$ scale uncertainty of the NNLO$_{\rm FTapprox}$ prediction~\cite{H:2026ksi}.
By including known effects in the central prediction rather than in the uncertainty budget, our results provide a smaller and physically motivated top-mass uncertainty and thereby provide a basis for precision studies of the Higgs self-interactions at the LHC.

\section*{Acknowledgements}
We are grateful to J.~Davies, G.~Heinrich, S.~Jones and K.~Sch\"onwald for their valuable comments on our manuscript. We thank E.~Brost for helpful discussions and careful reading of the manuscript and S.~O.~Moch for useful insights and advice on MSR implementation. 
S.E.J., R.V.K., and R.H.S. thank the Erwin-Schr\"odinger International Institute for Mathematics and Physics at the University of Vienna for partial support during the Programme ``New Paradigms for Harnessing Quantum Field Theory at Colliders'', July 27 - August 28, 2026, where parts of this work were completed.
C.D.P. and R.H.S. are supported by the U.S. Department of Energy under Contract No. DE-SC0012704.
S.E.J. is supported by the Swiss National Science Foundation Ambizione grant PZ00P2\_223524.
R.V.K. is supported by the Swiss High Energy Physics initiative for the FCC (CHEF) funded by SERI.

\bibliographystyle{elsarticle-num}
\bibliography{biblio}

\end{document}